\documentclass[12pt,preprint]{aastex}

\slugcomment{draft of \today}

\shorttitle{Nonthermal Radiation from ICM Shocks}
\shortauthors{Kang, Ryu, and Jones}

\def\ie{{\it i.e.,\ }}
\def\kms{~{\rm km\ s^{-1}}}
\def\yrs{~{\rm yrs}}
\def\muG{~{\mu\rm G}}
\def\cm3{~{\rm cm^{-3}}}

\begin{document}
\title{Diffusive Shock Acceleration Simulations of Radio Relics}

\author{Hyesung Kang$^1$, Dongsu Ryu$^2$\altaffilmark{,4}, and T. W. Jones$^3$}

\affil{$^1$Department of Earth Sciences, Pusan National University, Pusan 609-735, Korea: kang@uju.es.pusan.ac.kr\\
$^2$Department of Astronomy and Space Science, Chungnam National University, Daejeon 305-764, Korea: ryu@canopus.cnu.ac.kr\\
$^3$School of Physics and Astronomy, University of Minnesota, Minneapolis, MN 55455, USA: twj@msi.umn.edu}
\altaffiltext{4}{Corresponding Author}

\begin{abstract}

Recent radio observations have identified a class of structures, so-called radio relics, in clusters of galaxies. The radio emission from these sources is interpreted as synchrotron radiation from GeV electrons gyrating in $\mu$G-level magnetic fields. 
Radio relics, located mostly in the outskirts of clusters, seem to associate with shock waves, especially those developed during mergers. 
In fact, they seem to be good structures to identify and probe such shocks in intracluster media (ICMs), provided we understand the electron acceleration and re-acceleration at those shocks. 
In this paper, we describe time-dependent simulations for diffusive shock acceleration at weak shocks that are expected to be found in ICMs. 
Freshly injected as well as pre-existing populations of cosmic-ray (CR) electrons are considered, and energy losses via synchrotron and inverse Compton are included. 
We then compare the synchrotron flux and spectral distributions estimated from the simulations with those in two well-observed radio relics in CIZA J2242.8+5301 and ZwCl0008.8+5215. 
Considering that the CR electron injection is rather inefficient at weak shocks with Mach number $M \la$ a few,
the existence of radio relics could indicate the pre-existing population of low-energy CR electrons in ICMs.  
The implication of our results on the merger shock scenario of radio relics is discussed.

\end{abstract}

\keywords{acceleration of particles --- cosmic rays ---
galaxies: clusters: general --- shock waves}

\section{INTRODUCTION}

The presence of energetic nonthermal particles, especially electrons, in clusters of galaxies has been inferred from observations of so-called ``radio halos'' and ``radio relics'' \citep[see, e.g.,][for reviews]{ct02,gf04,ferr08,brug11}. The radio emission from these sources is interpreted as synchrotron radiation of cosmic-ray (CR) electrons. The radio halos center roughly in cluster cores and have low surface brightness with steep radio spectrum and low polarization. Radio relics, on the contrary, are isolated structures, typically located in the cluster outskirts but within virial radii. They often exhibit sharp edges, and most of them show strong polarization. In fact, with occasional pairings found in the opposite side of clusters and elongated morphologies, radio relics are commonly thought to reveal shock waves in intracluster media (ICMs) produced during mergers \citep[e.g.,][]{enss98,roett99,min01}. Unfortunately, relics are found mostly too far from cluster cores for their X-ray signatures to be easily detected. So only in a few cases, their association with ICM shocks have been established by X-ray observations \citep[e.g.,][]{fsnw10,aka11}. More than 40 relics have been identified in radio observations so far \citep[][and references therein]{nuza12}. Based on the spatial distribution of shocks seen in cluster formation simulations, it is predicted that coming radio surveys will easily identify hundreds more \citep[e.g.,][]{skill11,vazza12,nuza12}.

The observed synchrotron radiation is expected to come from CR electrons with Lorentz factors $\gamma_e \ga 10^4$, spiraling in $\sim \mu$G magnetic fields. The cooling time scale of such CR electrons due to synchrotron emission and inverse Compton (IC) scattering does not much exceed $\sim 10^8$ yrs (see equation (\ref{trad})). Advection or diffusion over that time would typically be limited to $\la 100$ kpc. So, the electrons have very likely been injected, accelerated or re-accelerated close to where they are seen in emission. 

Shocks, believed to be associated to observed radio relics, are obvious candidates for the acceleration or re-acceleration of the CR electrons. Suprathermal particles are known to be produced as an inevitable consequence of the formation of collisionless shocks in tenuous plasmas \citep[e.g.,][]{garat12}. If postshock suprathermal particles have sufficient rigidity to recross the shock transition, they can be further accelerated to become CRs through so-called Diffusive Shock Acceleration (DSA) \citep{bell78, dru83,maldru01}. Only a very small fraction of inflowing plasma particles are ``injected'' from the thermal pool into the CR population. Yet, in strong shocks, a sufficient number of CRs reach high energies, so that they extract a substantial fraction of the dissipated energy, allowing DSA to be efficient.

Shock waves are indeed common in the intergalactic space \citep[e.g.,][]{miniati00, ryuetal03}. They are induced by the supersonic flow motions produced during the hierarchical formation of the large-scale structure (LSS) in the universe. Those shocks are, in fact, the dominant means to dissipate the gravitational energy which is released during the LSS formation. They broadly reflect the dynamics of baryonic matter in the LSS of the universe, and, indirectly, dark matter. Simulations suggest that while very strong shocks form in relatively cooler environments in filaments and outside cluster virial radii, shocks produced by mergers and flow motions in hotter ICMs are relatively weak with Mach number $M \la$ a few \citep{ryuetal03,psej06,kangetal07,skillman08,hoeft08,vazza09,brug11}.

At weak shocks, however, DSA should be inefficient. This is expected from the fact that the particle energy spectrum associated with DSA is steep when the density compression across a shock is small. Also the relative difference between the postshock thermal and flow speeds is greater in weaker shocks. Consequently, the injection from thermal to nonthermal particles should be inefficient at weak shocks \citep[e.g.,][]{kangetal07}. At shocks with $M \la$ a few, much less than $\sim 10^{-3}$ of protons are thought to be injected into CRs and much less than $\sim 1$ \% of the shock ram pressure be converted into the downstream pressure of CR protons \citep{kr10}. For reference, recent Fermi observations of $\gamma$-ray emission from galaxy clusters, searching for $\gamma$-ray by-products of $p-p$ collisions, limit the pressure due to CR protons to less than $\sim 10$ \% of the gas thermal pressure there \citep{abdo10,ddcb10}. IACT (Imaging Atmospheric Cherenkov Technique) observations of TeV $\gamma$-ray suggest even a lower limit of $\la 1-2$ \% in core regions of some clusters \citep{alak12,arlen12}.

Injection and acceleration of electrons are even more problematic at weak shocks. Relativistic electrons and protons of the same energy are accelerated the same in DSA, since they have the same rigidity. But nonrelativistic electrons of a given energy have substantially smaller rigidities than protons, making them much harder to be injected at shocks from the thermal pool. As a consequence, the number of electrons injected and accelerated to the CR population is likely to be significantly smaller than that of CR protons, and so is the pressure of CR electrons at weak shocks.

Hot ICMs, on the other hand, should have  gone first through accretion shocks of high Mach numbers around clusters and filaments and then through weaker shocks inside those nonlinear structures \citep{ryuetal03, kangetal07}. Hence, it is expected that ICMs contain some CR populations produced through DSA at the structure formation shocks. In addition, in ICMs, nonthermal particles can be produced via turbulent re-acceleration \citep[e.g.,][]{bl07,bl11}. Moreover, secondary CR electrons are also continuously generated through $p-p$ collisions of CR protons with thermal protons of ICMs \citep[e.g.,][]{min01,pe04}. If radio relics form in media with such ``pre-existing'' CRs, the problem of inefficient injection at weak shocks can be alleviated.

In this paper, we study DSA of CR electrons at shocks expected to be found in ICMs, with and without pre-existing CR electrons. Since the shocks are mostly weak, the CR pressure is likely to be a small fraction of the thermal pressure \citep[see][]{kr11}. So we apply DSA in the test-particle regime. In the time-asymptotic limit without radiative losses, the test-particle DSA theory predicts a steady-state distribution of power-law for downstream CR electrons, $f_{e,2}(p)\propto p^{-q}$ with $q =3 \sigma/(\sigma-1)$, where $\sigma$ is the density compression ratio across a shock, when no pre-existing CR is assumed \citep{dru83}. If pre-existing CR electrons of a power-law distribution, $f_{e,1} \propto p^{-s}$, are assumed, the distribution of re-accelerated electrons approaches $f_{e,2}(p)\propto p^{-r}$ with $r = \min(q,s)$ at large momenta \citep[][and also see equation (\ref{f2})]{kr11}. The power-law distributions of $f_{e,2}(p)$ translate into the synchrotron/IC spectra of $j_{\nu} \propto \nu^{-\alpha}$ with $\alpha = (q-3)/2$ or $(r-3)/2$ \citep[e.g.,][]{za07, blasi10, kang11}. These properties provide essential benchmarks for expected spectral properties

We perform ``time-dependent'', DSA simulations of CR electrons for plane-parallel shocks, which include the energy losses due to  synchrotron and IC processes. Using the simulation data, we calculate the synchrotron emission from CR electrons, and model the synchrotron flux and spectral distributions from spherical shocks. We then compare the resulting distributions to those of two well-observed radio relics in clusters CIZA J2242.8+5301 \citep{vanweeren10} and ZwCl0008.8+5215 \citep{vanweeren11} in details.

The relic in CIZA J2242.8+5301 at redshift $z = 0.1921$ perhaps demonstrates the best evidence for DSA at merger shocks. It is located at a distance of $\sim 1.5$ Mpc from the cluster center and spans $\sim 2$ Mpc in length. The relic shows a spectral index gradient towards the cluster center. The spectral index, measured between 2.3 and 0.61 GHz, steepens from $-0.6$ to $-2.0$ across the relic. The relic is strongly polarized at the $50 - 60$ \% level, indicating ordered magnetic fields aligned with the relic. In the opposite, southern part of the cluster, an accompanying fainter and smaller relic is found. The relic in ZwCl0008.8+5215 at $z = 0.1032$ is found at a distance of $\sim 0.85$ Mpc from the cluster center and has a linear extension of $\sim 1.4$ Mpc. It also shows the steepening of the spectral index towards the cluster center. The spectral index, measured between 1382 and 241 MHz, changes from $-1.2$ to $-2.0$ across the relic. The polarization fraction is less with $\la 25$ \%. It also has an accompanying relic of a linear extension of $\sim 290$ kpc in the opposite, western side of the cluster.

In Section 2 we describe the numerical method and the models for magnetic field, diffusion, electron injection, and pre-existing CR electron population. We present analytic evaluations for some features in the CR electron energy spectrum and synchrotron emission spectrum in Section 3. The results of simulations are presented and compared with observations of the previously mentioned radio relics in Section 4. Summary follows in Section 5.

\section{DSA SIMULATIONS OF CR ELECTRONS}

\subsection{Numerical Method}

We simulate DSA of CR electrons at gasdynamical shocks in one-dimensional plane-parallel geometry. Shocks in ICMs, especially merger shocks, are expected to persist over $\ga 10^9$ yrs, a substantial fraction of the cluster lifetime \citep[e.g.,][]{skill11}. On the other hand, the time scales over which electrons are accelerated and cool are much shorter (see Eq. [\ref{trad}] below). So we assume that the shock structure remains steady. Assuming that the CR feedback to the flow is negligible at weak shocks in the test-particle limit, the background flow, $u$, is given by the usual shock jump condition. Then, the time-dependent evolution of the CR electron distribution, $f_e(t,x,p)$, which is averaged over pitch angles, can be followed by the following diffusion convection equation,
\begin{equation}
{\partial g_e\over \partial t} + u {\partial g_e \over \partial x} = {1\over{3}} {\partial u \over \partial x} \left( {\partial g_e\over \partial y} -4g_e \right) + {\partial \over \partial x} \left[\kappa(x,y) {\partial g_e \over \partial x}\right] + p {\partial \over {\partial y}} \left( {b_e\over p^2} g_e \right)
\label{dce},
\end{equation}
where $g_e=p^4 f_e$, $y=\ln(p/m_e c)$, $m_e$ is the electron mass, $c$ is the speed of light, and $\kappa(x,y)$ is the spatial diffusion coefficient \citep{skill75}.

Here, $b_e(p) = (4 e^4/ 9 m_e^4 c^6) B_{\rm eff}^2 p^2$ represents the cooling of CR electrons due synchrotron and IC losses in cgs units, where $e$ is the electron charge.  The ``effective'' magnetic field strength, $B_{\rm eff}^2 \equiv B^2 + B_{\rm CBR}^2$, includes the equivalent strength of the cosmic background radiation with $B_{\rm CBR}=3.24\muG(1+z)^2$ at redshift $z$. The cooling time scale for electrons is given as
\begin{equation}
t_{\rm rad} (\gamma_e) = { p \over {b_e(p)}} = 9.8\times 10^{7} \yrs \left({B_{\rm eff} \over {5 \muG}}\right)^{-2} \left({\gamma_e \over 10^4 }\right)^{-1},
\label{trad}
\end{equation}
where $\gamma_e$ is the Lorentz factor of CR electrons.

The equation in (\ref{dce}) is solved using the test-particle version of the CRASH (Cosmic-Ray Amr SHock) code \citep[see][for details]{kej11}.

\subsection{Models for Magnetic Field and Diffusion}

Here, shocks are assumed to be gasdynamical for simplicity, that is, magnetic fields do not modify the background flow of the shock. In ICMs, magnetic fields of an inferred strength of order $\mu$G \citep{ct02,gf04} are dynamically unimportant, since their energy density is less than $\sim 10\ \%$ of the thermal energy density \citep[e.g.,][]{ryuetal08}. However, magnetic fields, especially in the downstream region, are the key that governs DSA and the synchrotron cooling and emission of CR electrons. Theoretical studies have shown that efficient magnetic field amplification via resonant and non-resonant wave-particle interactions is an integral part of DSA at strong shocks \citep{lucek00, bell04}. In addition, magnetic fields can be amplified by turbulent motions behind shocks \citep{giacal07,inoue09}. Yet, these plasma processes are complex and their roles are not yet entirely certain, especially at weak shocks. So here we adopt a simple model in which the magnetic field strength is amplified by a constant factor of $\chi$ across the shock, that is, $B_2= \chi B_1$. Hereafter, we use the subscripts `1', and `2' to label conditions in the preshock and postshock regions, respectively.

For $\kappa$, we adopt a Bohm-like diffusion coefficient with weaker non-relativistic momentum dependence,
\begin{equation}
\kappa(x,p) = \kappa^* \cdot \left({p \over {m_e c}} \right), 
\label{kappa}
\end{equation}
where $\kappa_1^*= m_e c^3/(3eB_1)=1.7\times 10^{19} {\rm~ cm^2~ s^{-1}} (B_1/1\muG)^{-1}$ in the preshock region and $\kappa_2^*=\kappa_1^*/\chi$ in the postshock region.

\subsection{Injection of Electrons}

As pointed in Introduction, the injection of electrons is expected to be much harder than that of protons in the so-called thermal leakage injection model. Because complex plasma interactions among CRs, waves, and the underlying gas flow are not fully understood, it is not yet possible to predict from first principles how particles are injected into the first-order Fermi process \citep[e.g.,][]{maldru01,garat12}. In addition, postshock thermal electrons, which have gyro-radii smaller than those of thermal protons, need to be pre-accelerated to several times the peak momentum of thermal protons, $p_{\rm p,th}$, before they can re-cross the shock transition layer. Here, $p_{\rm p,th}= \sqrt{2m_p k_B T_2}$, where $T_2$ is the postshock gas temperature and $k_B$ is the Boltzmann constant. Recently several authors have suggested pre-acceleration mechanisms based on plasma interactions with fluctuating magnetic fields that are locally quasi-perpendicular to the shock surface \citep[e.g.][]{burgess06,amano09,guo10,riq11}. But the detailed picture of the electron injection is not well constrained by plasma physics. Observationally, the ratio of CR electron number to proton number, $K_{e/p} \sim 0.01$, is commonly inferred for strong supernova remnant shocks, since about 1\% of the Galactic CR flux near a GeV is due to electrons \citep{reynolds08}. But this ratio is rather uncertain for weak shocks under consideration.

So here we adopt a simple model in which the postshock electrons above a certain injection momentum, $p_{\rm inj}= Q_{\rm inj} p_{\rm p,th}$, are assumed to be injected to the CR population. Here, $Q_{\rm inj}$ is a parameter that depends on the shock Mach number and turbulent magnetic field amplitude in the thermal leakage injection model \citep{kr10}. The CR electron number density or, equivalently, the distribution function at $p_{\rm inj}$ at the shock location $x_s$, $f_e(x_s,p_{\rm inj})$, will be scaled to match the observed flux of radio relics (see Sections 3.1 and 4.2).

\subsection{Pre-existing CR Electrons}

We consider the population of pre-existing CR electrons, along with that of freshly injected electrons at the shock. However, the nature of pre-existing CR electrons in ICMs is not well constrained. If they were generated at previous, external and internal shocks, a spectral slope of $s \sim 4.0-5.3$ is expected for $M \ga 2$, close to the acceleration site. However, since their lifetime in equation (\ref{trad}) is much shorter than that of host clusters, it is unlikely that they are directly responsible for the pre-existing electron population we consider here. Any pre-existing CR electron should be locally produced, possibly either through $p-p$ collisions of CR protons with thermal protons or via turbulent re-acceleration of some populations (possibly including $p-p$ secondary electrons), as noted in Introduction. \citet{pet08} have shown that turbulent injection of CR electrons from the thermal pool in ICMs is unlikely. The slope of protons re-accelerated by turbulence should be close to $s \sim 4$ \citep[see, e.g.,][]{chan05}, but that of electrons is strongly modified by coolings \citep{bl07,bl11}. The slope of secondary electrons from $p - p$ collisions would be roughly $s \sim 4/3(s_p - 1)$ \citep{ms94}, where $s_p$ is the slope of CR protons, so typically, $s \sim 4 - 6$.
In summary, pre-existing CR electrons may contain many different populations with different degrees of radiative cooling and may not be represented by a single power-law. 

For simplicity, here we adopt a power-law form,
\begin{equation}
f_{e,1}(p)=f_{\rm pre}\cdot \left(p \over p_{\rm inj}\right)^{-s},
\label{f1}
\end{equation}
with slope $s$, as the model spectrum for pre-existing CR electrons. In modeling of specific radio relics, the value of $s$ will be chosen as $s= 2\alpha_{\rm obs}+3$, where $\alpha_{\rm obs}$ is the observed mean spectral index. The amplitude, $f_{\rm pre}$, is set by the ratio of upstream CR electron pressure to gas pressure, $R_1\equiv P_{\rm CRe,1}/P_{\rm g,1}$. Here, $R_1$ is a parameter that will be scaled to match the observed fluxes of radio relics (see Sections 3.1 and 4.2).

\section{ANALYTIC EVALUATIONS} 

We first consider some features in the CR electron energy spectrum and synchrotron emission spectrum for plane-parallel shocks, to provide analytic estimations for the simulation results presented in the next section.

\subsection{Basic Features in CR Electron Spectrum}

In the test-particle regime of DSA, the distribution of freshly injected and accelerated electrons at the ``shock location'' can be approximated, once it reaches equilibrium, by a power-law spectrum with super-exponential cutoff, 
\begin{equation}
f_{e,2}(p) \approx f_{\rm inj}\cdot \left(p \over p_{\rm inj} \right) ^{-q} \exp\left(-{p^2 \over p_{\rm eq}^2} \right), 
\label{f2o}
\end{equation}
where $q = 3 \sigma /(\sigma -1 )$ \citep{kang11}. In the case that $B_2 = \sigma B_1$, that is, the jump in the magnetic field strength across the shock is assumed to be same as the density jump, $\chi=\sigma$, and $\kappa_2=\kappa_1/\sigma$, the cutoff momentum, which represents the balance between DSA and the radiative cooling, becomes 
\begin{equation}
p_{\rm eq}= {m_e^2 c^2 u_s \over \sqrt{4e^3q/27}} \left({B_1 \over {B_{\rm eff,1}^2 + B_{\rm eff,2}^2}}\right)^{1/2}.
\label{peq}
\end{equation} 
The corresponding Lorentz factor for typical merger shock parameters is then 
\begin{equation}
\gamma_{e, {\rm eq}} \approx 2\times 10^9 \ q^{-1/2} \left({u_s \over {3000 \kms}}\right) \left({B_1 \over {B_{\rm eff,1}^2 + B_{\rm eff,2}^2}}\right)^{1/2}.
\label{gammaeq}
\end{equation}
Hereafter, the magnetic field strength is given in units of $\mu$G. The acceleration time for electrons to reach $p_{\rm eq}$, so the time for the equilibrium to be achieved, is estimated as
\begin{equation}
t_{\rm eq}\approx (2.4\times 10^4 \yrs) \ q^{1/2} B_1^{-1/2} (B_{\rm eff,1}^2 + B_{\rm eff,2}^2)^{-1/2} \left({u_s \over {3000 \kms}}\right)^{-1}.
\label{teq}
\end{equation}
This is much shorter than the typical time scale of merger shocks, $\ga 10^9 \yrs$. For $t \ga t_{\rm eq}$, the DSA gains balance the radiative losses and the electron spectrum near the shock location asymptotes to a steady-state \citep{kang11}.

With pre-existing CR electrons given in equation (\ref{f1}), the electrons distribution at the shock location can be written as the sum of the pre-existing/re-accelerated and freshly injected/accelerated populations,
\begin{eqnarray}
f_{e,2}(p) \approx \left\{ \begin{array}{cc}
\left[ {q \over (q-s)} \left(1-\left({p \over p_{\rm inj}}\right)^{-q+s}\right) f_{\rm pre} \left({p\over p_{\rm inj}}\right)^{-s} + f_{\rm inj} \left({p\over p_{\rm inj}}\right)^{-q} \right] \exp\left(-{p^2 \over p_{\rm eq}^2} \right), &{\rm when}\ s \neq q\\ 
\left[ s \ln \left({p\over p_{\rm inj}}\right) f_{\rm pre} \left({p\over p_{\rm inj}}\right)^{-s} + f_{\rm inj} \left({p\over p_{\rm inj}}\right)^{-q} \right] \exp\left(-{p^2 \over p_{\rm eq}^2} \right), &{\rm when}\ s = q.
\end{array} \right.
\label{f2}
\end{eqnarray}
\citep{kr11}. The relative importance of pre-existing to freshly injected populations depends on $f_{\rm pre}$ and $f_{\rm inj}$, as well as on the slopes $s$ and $q$ in our model.
For the sake of convenience, hereafter we will use the term ``injected'' electrons for those injected at the shock and then accelerated by DSA and the term ``re-accelerated'' electrons for those accelerated from the pre-existing population.

We here define the CR electron number fraction, $\xi_e \equiv {n_{CRe,2}/n_{e,2} }$, as the ratio of CR electron number to thermal electron number in the postshock region. Here $n_{CRe,2}$ includes CR electrons accelerated from both the pre-existing and freshly injected populations. Considering that the CR proton number fraction is likely to be $\xi_p \la 10^{-4}$ at weak shocks \citep{kr10} and $K_{e/p} \sim 0.01$, $\xi_{e} \sim 10^{-6}$ could be regarded as a canonical value. We note that the resulting radio emission is linearly scaled with both $\xi_e$ and the preshock gas density, $n_1$, in the test-particle regime, so the combined parameter, $n_1 \xi_e$, can be treated as a free parameter. We here fix the preshock gas density, $n_1 = 10^{-4} {\rm cm^{-3}}$, as a fiducial parameter, but vary $\xi_e$ to match the observed fluxes of radio relics. Another measure is the ratio of postshock CR electron pressure to gas pressure, $R_2= P_{\rm CRe,2}/P_{\rm g,2}$, which depends on both $\xi_e$ and the slopes $q$ and $s$. In modeling of specific radio relics in Section 4.2, we will determine the set of values for $\xi_e$, $R_2$ and $R_1$, that matches the observed level of radio flux.

If we ignore for the moment the modest influence of continued DSA downstream of the shock, we can follow the electron population that advects downstream
by solving the following equation :
\begin{equation}
{d g_e\over d t}+ V \cdot {\partial g_e\over \partial y} = 0,
\label{gecool}
\end{equation}
where $d/dt \equiv \partial/\partial t + u \partial/\partial x$ and $V = - b_e(p)/p = -C e^y $. Here, $C = (4 e^4/ 9 m_e^4 c^6) B_{\rm eff}^2$ is a constant. This is basically the equation for downward advetion in the space of $y=\ln(p/m_ec)$ due to radiative cooling. The general solution of the equation is
\begin{equation}
g_e(p,t)= G(e^{-y}-Ct) = G\left(p \over 1-t/t_{\rm rad}\right),
\label{gapprox}
\end{equation}
where $t_{\rm rad} = 1/Ce^y$ is the electron cooling time scale. This provides the approximate distribution of CR electrons at the distance $d = u_2 t$ downstream from the shock, where $u_2$ is the downstream flow speed.

For instance, if the distribution function of the ``injected'' electrons at the shock location ($d=0$) is the power-law spectrum,
 $g_e(p,0) =g_{\rm inj} (p/p_{\rm inj})^{-q+4}$, the downstream spectrum can be approximated as
\begin{equation}
g_e(p,d) =g_{\rm inj} \left[ p \over{ (1- d/u_2t_{\rm rad})p_{\rm inj}}\right]^{-q+4}.
\label{ganal}
\end{equation}
It should be straightforward to apply the same approximation to the full spectrum given in equation (\ref{f2}). In Figures 1 and 2, we compare the distributions described by equation (\ref{gapprox}) with those from time-dependent DSA simulations, demonstrating that equation (\ref{gapprox}) provides reasonable approximations to the solutions of full DSA simulations (see Table 1 for specific model parameters).

\subsection{Basic Features in Synchrotron Emission Spectrum}

Since the synchrotron emission from mono-energetic electrons with $\gamma_{\rm e}$ has a broad peak around
$\nu_{\rm peak} \approx 0.3 (3eB/4\pi{m_e c}) \gamma_e^2$, for a given observation frequency, $\nu_{\rm obs}$, the greatest contribution comes from electrons of the Lorentz factor,
\begin{equation}
\gamma_{\rm e,peak} \approx 1.26\times 10^4 \left({ \nu_{\rm obs} \over {1 {\rm GHz}}}\right)^{1/2} \left({B \over 5\muG} \right)^{-1/2} (1+z)^{1/2}.
\label{fpeak}
\end{equation}
Using equations (\ref{trad}) and (\ref{fpeak}), the cooling time of the electrons emitting at $\nu_{\rm obs}$ can be estimated approximately as 
\begin{equation}
t_{\rm rad} \approx 8.7\times 10^8 \yrs \left({{B_2^{1/2}} \over {B_{\rm eff,2}^2}}\right) \left({ \nu_{\rm obs} \over {1 {\rm GHz}}}\right)^{-1/2} (1+z)^{-1/2}.
\end{equation}
The cooling length behind the shock, $u_2 t_{\rm rad}$, then becomes 
\begin{equation}
L_{\rm rad} \approx 890 {\rm kpc} \left( {u_2 \over {10^3 \kms}} \right) \left({ {B_2^{1/2}} \over {B_{\rm eff,2}^2}}\right) \left({ \nu_{\rm obs} \over {1 {\rm GHz}}}\right)^{-1/2} (1+z)^{-1/2}.
\label{lrad}
\end{equation}
Note that $B_{\rm eff,2}^2/B_2^{1/2}\sim 15-25$ for the model parameters considered here.
Again, $t_{\rm rad}$ is shorter than the typical time scale of merger shocks, $\ga 10^9 \yrs$. So $L_{\rm rad}$ should represent the width of radio emitting region at $\nu_{\rm obs}$ behind plane-parallel shocks. In radio relics, however, the observed width is constrained by both $L_{\rm rad}$ and the projection angle of spherical shocks (see Section 4.2).

The cutoff energy in the electron spectrum due to the radiative cooling decreases linearly with the distance from the shock location, that is, $\gamma_{e,{\rm cut}} \propto d^{-1}$, as expected from equation (\ref{trad}) and shown in Figure 1. At the farthest downstream point, $d=u_2 t$, where $t$ is the shock age, the cutoff energy becomes 
\begin{equation}
\gamma_{e,{\rm br}}(t) \approx 9.82 \times 10^2 \left({t \over 10^9 \yrs}\right)^{-1} \left({B_{\rm eff,2} \over {5 \muG}}\right)^{-2}.
\label{pbr}
\end{equation}
If the electron distribution function at the shock location has a power-law form, $n_e(x_s,\gamma_e)\propto \gamma_e^{-r}$, then the volume-integrated electron spectrum downstream steepens by the power-law index of one, \ie $N_{e,2} \propto \gamma_e^{-(r+1)}$ for $\gamma_e > \gamma_{e,{\rm br}}$. It is because the width of the spatial distribution of electrons with $\gamma_e$ decreases as $\gamma_e^{-1}$ \citep{za07, kang11}. As a consequence, the ``volume-integrated'' synchrotron spectrum from aged electrons has a spectral break, \ie an increase of the spectral index $\alpha$ by $+0.5$, at
\begin{equation}
\nu_{\rm br} = 0.3 \frac{3}{4\pi} { {eB_2} \over {m_e c}} \gamma_{e,{\rm br}}^2 \approx 6.1\times 10^6{\rm Hz}\left({t \over 10^9 \yrs}\right)^{-2} \left({B_2 \over {5 \muG}}\right) \left({B_{\rm eff,2} \over {5 \muG}}\right)^{-4}.
\label{fbr}
\end{equation}
So the shock age may be estimated from the break frequency $\nu_{\rm br}$, if the magnetic field strength is known.

\section{RESULTS OF DSA SIMULATIONS}

\subsection{Plane-Parallel Shocks}

The model parameters of our simulations for plane-parallel shocks are summarized in Table 1. Here, $z$ is the redshift, $c_{\rm s,1}$ is the preshock sound speed, $M$ is the shock Mach number, $u_2$ is the postshock flow speed, $s$ is the power-law slope of pre-existing CR electrons, and $B_2$ is the postshock magnetic field strength. The model name in the first column includes the values of $M$, $B_2$, and $s$; for models without pre-existing CRs, ``I'' (injection only) is specified. For instance, M4.5B7I stands for the model with $M=4.5$, $B_2=7\muG$, and injected CR electrons only (no pre-existing CRs), while M2B2.3S4.2 stands for the model with $M=2.0$, $B_2=2.3\muG$, and $s=4.2$. For the preshock magnetic field strength, $B_1=1 \muG$ is adopted for all models, which is close to the typical quoted value in cluster outskirts \citep[see][and references therein]{brug11}. Once $B_1 < B_{\rm CBR}$, the IC cooling dominates, and the exact value of $B_1$ is not important in our models. The model parameters are chosen to match the observed properties of radio relics in clusters CIZA J2242.8+5301 and ZwCl 0008.8+5215 (see the next subsection for details). For example, $M=4.5$ or $s=4.2$ is chosen to match the observed spectral index, $\alpha=0.6$, of the relic in CIZA J2242.8+5301, and $M=2$ or $s=5.4$ is chosen to match $\alpha=1.2$ of the relic in ZwCl 0008.8+5215. For reference, the shock compression ratio is $\sigma=3.5$ for $M=4.5$ and $\sigma=2.3$ for $M=2$. The values of $u_2$ and $B_2$ are chosen to match the observed width of the relics, since they determine the cooling length as shown in equation (\ref{lrad}).

Figure 1 shows the CR electron distribution at different locations downstream of the shock, after it has reached the steady state, for M4.5B3.5I, M2B7S4.2, M2B2.3I and M2B2.3S5.4 models. For the comparison of different models, here the postshock CR electron number fraction is set to be $\xi_e = 10^{-6}$, which sets the vertical amplitude. 
The injection-only models exhibit the power-law distributions with cutoffs due to the cooling, as discussed in the previous section. 
In M2B7S4.2 model, the electrons accelerated from the injected population are important only at low energies ($\gamma_e \la 10^{2.5}$)
and they dominate in terms of particle number, because the ``injected'' spectrum is much softer than the ``re-accelerated'' spectrum (i.e. $q>s$). 
The electrons accelerated from the pre-existing population, on the other hand, dominate at higher energies including $\gamma_e \sim 10^4$ and they are most relevant for the synchrotron emission at $\nu \sim 1$ GHz.  The slope of the accelerated spectrum at high energies is similar to that of the pre-existing spectrum,
which is consistent with equation (\ref{f2}). 
On the contrary, in M2B2.3S5.4 model with $s\approx q$, the ``injected'' electrons are negligible even at low energies. This difference comes about, because with similar numbers of pre-existing CRs, the amplitude $f_{\rm pre}$ is larger in M2B2.3S5.4 (with $s=5.4$) than in M2B7S4.2 (with $s=4.2$). The numbers of injected electrons should be similar in the two models, because the shock Mach number is the same. 
 Note that the re-accelerated spectrum flattens by a factor of $\ln(p)$, as shown in equation (\ref{f2}), because $s \approx q$ in this model.

The left column of Figure 2 shows the spatial profile of the electron distribution function, $g_e(\gamma_e,x)$, at two specific energies ($\gamma_e$) as a function of the downstream distance for the M4.5B7I, M4.5B3.5I and M2B7S4.2 models. For each model the Lorentz factors are calculated for $\nu_{\rm obs}$ = 0.61 GHz and 2.3 GHZ according to equation (\ref{fpeak}). The upper/lower curves represent $g_e$ of the lower/higher values of $\gamma_e$, respectively. The right column of Figure 2 shows the synchrotron emission, $j_{\nu}(x)$, at $\nu_{\rm obs}$ = 0.61 GHz (upper curves) and 2.3 GHZ (lower curves). The solid lines show $g_e$ and $j_{\nu}$ calculated from the DSA simulation results, while the dashed lines show the approximate solutions calculated with equation (\ref{gapprox}). The figure demonstrates that the lower energy elections advect further from the shock before cooling than higher energy electrons, so the lower-frequency radio emission has larger widths than the higher-frequency one. According to equation (\ref{lrad}), the cooling lengths of the electrons emitting at 0.61 and 2.3 GHz are $L_{\rm rad} \approx $ 40 and 20 kpc, respectively, in the three models.

\subsection{Modeling of Radio Relics}

As noted above, it should be sufficient to employ the plane shock approximation to compute the distributions of CR electrons and their emissivities as a function of the distance from the shock surface.
In observed radio relics, however, radio emitting shells are likely to be curved with finite curvatures along the observer's line of sight (LoS) as well as in the plane of the sky. So in modeling of radio relics, the curved shell needs to be projected onto the plane of the sky. In that case LoS's from the observer will transect a range of shock displacements, and this needs to be taken into account when computing the observed brightness distribution of model relics. Following the approach of \citet{vanweeren10,vanweeren11}, we consider a piece of a spherical shell with outer radius $R_s$, subtended along the LoS from $+\psi$ to $-\psi$ so for the total angle of $2 \psi$. Then, $R_s$ and the projection angle $\psi$ are the parameters that fix the shape of the curved shell to be projected onto the plane of the sky. The synchrotron emissivity, $j_{\nu}\ ({\rm erg~cm^{-3}~s^{-1}~Hz^{-1}~str^{-1}})$, at each point behind the curved shock is approximated as that downstream of plane-parallel shocks discussed in the previous subsection. Since we do not consider the polarization of synchrotron emissions here, so, for simplicity, the magnetic field lines are assumed to lie in the plane of the sky; that is, the angle between the magnetic field vectors and the LoS is fixed at $90^\circ$.

The synchrotron intensity is calculated by integrating the emissivity along the LoS, $I_{\nu}(r)= \int j_{\nu} d {\it l}\ ({\rm erg~cm^{-2}~s^{-1}~Hz^{-1}~str^{-1}})$, where $r$ is the distance behind the projected shock edge in the plane of the sky. The bound of the path length, $l$, for given $r$ is determined by $R_s$ and $\psi$. Then, the observed flux is estimated, assuming a Gaussian beam with e-width, $\theta$, as
\begin{equation}
S_{\nu}(r) \approx I_{\nu}(r) \pi \theta^2 (1+z)^{-3},
\label{Snu}
\end{equation}
where $\nu = \nu_{\rm obs}(1+z)$.

Figure 3 shows the profiles of the synchrotron flux, $S_{\nu}(r)$, at $\nu_{\rm obs} = 0.61$ GHz (left column) and the spectral index, $\alpha=-d\ln S_{\nu}/d\ln \nu$, estimated with the fluxes at $\nu_{\rm obs} = 0.61$ and 1.4 GHz (right column) for the M4.5B7I, M4.5B3.5I and M2B7S4.2 models, which are designed to reproduce the radio relic in CIZA J2242.8+5301. The flux is calculated with the beam of $\theta^2 = \theta_1 \theta_2/(4 \ln 2)$, $ \theta_1 \theta_2=16.7^{"} \times 12.7^{"}$. They are compared with the ``deconvolved'' profile of observed flux taken form Figure 4 of \citet{vanweeren10} (filled circles). Since the observed flux is given in an arbitrary unit in their paper, we scale it so that the peak value of $S_{\nu}(r)$ becomes 5 mJy, which is close to the observed value (private communication with R. J. van Weeren). The radius of the spherical shock is set to be $R_s=1.5$ Mpc and two values of projection angle, $\psi = 10^{\circ}$ and $20^{\circ}$, are considered. The observed profile is well fitted by the three models, if $\psi=10^{\circ}$ is taken. In M4.5B7I and M4.5B3.5I, different values of $u_2$ are assumed to match the observed width (see Table 1). The observed value of the spectral index at $r=0$, $\alpha=0.6$, is reproduced either in the injection-only models with $M=4.5$ or in the model with pre-existing CRs with the slope $s = 4.2$, as noted in the previous subsection.

For the fiducial preshock particle density of $n_1=10^{-4} \cm3$, the values of the postshock electron CR number fraction required to match the peak flux of 5 mJy are $\xi_e=7.6\times 10^{-8},~ 2.3\times 10^{-7}$, and $2.6\times 10^{-7}$ for M4.5B7I, M4.5B3.5I, and M2B7S4.2, respectively. In M2B7S4.2 model the ratio of the pressure of pre-existing CR electrons to gas pressure far upstream is $R_1\sim 6.7\times 10^{-5}$. Those values of $\xi_e$ and $R_1$ are modest enough that they probably are not in conflict with the values expected in clusters. Our results demonstrate that if the pre-existing electron population is considered, the radio relic in CIZA J2242.8+5301 can be reproduced even with weak shocks of $M \sim 2$ or so. We note that $R_1$ is a model parameter that sets the amplitude, $f_{\rm pre}$, of the upstream population,
while the fraction $\xi_e$ is the outcome of DSA of both pre-existing and injected electrons.
As noted in Figure 1, in M2B7S4.2 model the fraction $\xi_e$ is determined mostly by the ``injected'' population at low energies, 
while the radio emission is regulated mostly by the ``re-accelerated'' population at $\gamma_e \sim 10^4$. 
So we should obtain the similar radio flux even with a much lower injection rate for this model, and the resulting $\xi_e$ could be much smaller than the current value of $2.6\times 10^{-7}$.

We point that the radio relic in CIZA J2242.8+5301 is subtended in the plane of the sky over the angle of $\sim 60 - 70^\circ$. This means that the surface of the shock responsible for the relic should be highly elongated with the aspect ratio of $\sim (60 - 70^\circ) / (2 \psi) =\ \sim 3 - 3.5$ when $\psi=10^{\circ}$ is adopted. It would not be trivial, if not impossible, for such structure to be induced in merger events in clusters. Or the relic may actually consist of a number of substructures, which is hinted by the variations in the observed flux profile along the arc in the plane of the sky.
 
The left column of Figure 4 shows the synchrotron flux profiles at $\nu_{\rm obs} = 1.38$ GHz, for M2B2.3I and M2B2.3S5.4 models, which are designed to reproduce the radio relic in ZwCl008.8+5215. The flux is calculated with $\theta^2 = \theta_1 \theta_2/(4 \ln 2)$, $ \theta_1 \theta_2=23.5^{"} \times 17.0^{"}$. We note that this beam size is fine enough that the convolved profiles with a Gaussian beam (dotted and long-dashed lines) are very similar to the unconvolved profiles (solid lines). The profiles are compared with the observed profile given in Figure 16 of \citet{vanweeren11} (filled circles). Again the observed flux is given in an arbitrary unit, so it is scaled at 5 mJy at the peak. The right column shows the profiles of $\alpha$, estimated with fluxes at $\nu_{\rm obs} = 0.24$ and 1.38 GHz, along with the observed $\alpha$ also taken from Figure 16 of \citet{vanweeren11} (filled circles). The shock radius is assumed to be $R_s=1.0$ Mpc and two values of projection angle, $\psi = 25^{\circ}$ and $30^{\circ}$, are considered. The two models shown are the same except the existence of pre-existing CR electrons in M2B2.3S5.4 model. In M2B2.3S5.4, the ``re-accelerated'' population dominates over the ``injected'' population. Yet, the two models give similar profiles of $S_{\nu}$ and $\alpha$. We see that in our models $\psi=30^{\circ}$ gives good fits to the observed profiles of $S_{\nu}$ and $\alpha$, while \citet{vanweeren11} argued that $\psi=22^{\circ}$ seems to give a reasonable fit. Note that they adopted $u_2=750\kms$ and $B_2=2\muG$, giving $L_{\rm rad}=40$ kpc, while in our models $u_2=1100\kms$ and $B_2=2.4\muG$, giving $L_{\rm rad}=57$ kpc.

For the assumed value of $n_1=10^{-4} \cm3$, the postshock CR electron number fraction required to match the peak flux of 5 mJy is $\xi_e =2.1\times 10^{-4}$ for M2B2.3I, which is six times larger than $\xi_e =3.3\times 10^{-5}$ for M2B2.3S5.4. This is because the spectral shapes of CR electron spectrum below $\gamma_e \la 10^{2.5}$ are different in the two models (see Fig. 1 and discussion in the previous subsection). The number fraction of CR electrons for M2B2.3I seems too large, considering that the postshock proton CR number fraction is likely to be $\xi_p \la 10^{-4}$ for $M = 2$ \citep{kr10}. In M2B2.3S5.4, on the other hand, the ratio of upstream CR electrons pressure to gas pressure is $R_1\sim 1.2\times 10^{-3}$. This seems to be marginal, that is, not inconsistent with expected values, considering that the ratio of CR proton pressure to gas pressure is $\la 10^{-2} - 10^{-1}$ in ICMs as noted in Introduction. But we should point that the values of $\xi_e$ and $R_1$ in these two models are dominated by low-energy CR electrons with $\gamma_e \la 10^3$ (see Figure 1), which do not contribute much to the synchrotron radiation observed in radio relics. So if the  ``injected'' population in M2B2.3I consists of electrons with $\gamma_e \ga 10^3$ only, the required values of $\xi_e$ could be reduced by a factor of $\sim 10$, easing down the constraint. 
Since we do not understand fully the plasma interactions involved in the pre-acceleration and injection of electrons at the shock, the detailed spectral shape of those low energy electrons are very uncertain.

The top panels of Figure 5 show the profiles of the intensity, $I_{\nu}(r)= \int j_{\nu} d {\it l}$, at 6 cm (5 GHz), 20 cm (1.5 GHz), and 91 cm (0.33 GHz) in arbitrary units as a function of $r$ for the M4.5B3.5I, M2B7S4.2 and M2B2.3I models. Here, the projection angle is set to be $\psi=30^{\circ}$. Since the emissivity $j_{\nu}$ decreases downstream of the shock, while the path length increases with $r$, the profiles of $I_{\nu}$ exhibit non-monotonic behaviors. For example, the profiles at 6 cm show a slightly concave turnover before it decreases abruptly at $r \approx 200$ kpc. The middle panels show the spectral indices, $\alpha_{20}^6$ (solid lines) calculated between 6 and 20 cm and $\alpha_{91}^{20}$ (dashed lines) calculated between 20 and 91 cm, when the projection angle is set to be $\psi=10^{\circ}$, $20^{\circ}$ and $30^{\circ}$. The general trend is the increase of $\alpha_{20}^6$ and $\alpha_{91}^{20}$ as we move away from the projected shock edge at $r = 0$, reflecting the effects of radiative cooling. Also $\alpha_{20}^6 > \alpha_{91}^{20}$, that is, the slope is steeper at higher frequencies. The bottom panels show the color-color diagram of $\alpha_{91}^{20}$ versus $\alpha_{20}^{6}$. The rightmost point ($\alpha_{20}^6 = \alpha_{91}^{20} = \alpha_s$) corresponds to the projected shock edge. Away from the edge, the loci move towards the lower left direction. In both middle and bottom panels, the spectral slopes also show a slightly concave turnover for large projection angles of $\psi=20^{\circ}$ and $30^{\circ}$. Recently, \citet{vanweeren12} reported the color-color diagram for the so-called ``Toothbrush'' relic in cluster 1RXS J0603.3+4214, which shows a spectral behavior that is consistent with the cooled electron population downstream of the shock. 

\section{SUMMARY}

In an effort to refine our understandings of radio relics in clusters of galaxies, we have performed time-dependent, DSA simulations of CR electrons and calculated the synchrotron emission from CR electrons for plane-parallel shocks. The energy losses due to synchrotron and IC have been explicitly included. Weak shocks expected to be found in ICMs have been considered. Both the cases with and without pre-existing CR electrons have been considered. The relevant physics of DSA and cooling is well represented by plane-parallel shocks, since the time scales over which electrons are accelerated and cool are much shorter than the lifetime of merger shocks in clusters and the radio emission is confined to a region of small width behind the shock front. We then have modeled the synchrotron flux and spectral distributions from spherical shocks by approximating them with plane-parallel shocks and projecting to the plane of the sky for the angle from $+\psi$ to $-\psi$ along the LoS. For the specific models which are designed to reproduce radio relics in clusters CIZA J2242.8+5301 and ZwCl0008.8+5215, we have compared the resulting distributions with observed ones in details.

The main results are summarized as follows:

1) The CR electron spectrum becomes steady, after the DSA gains balance the radiative losses. The spectrum at the shock location is well approximated by a distribution with super-exponential cutoff at $p_{\rm eq}$, $f_{e,2}(p) \propto \exp(-p^2/p_{\rm eq}^2)$. The full expressions of $f_{e,2}(p)$ and $p_{\rm eq}$ are given in equations (\ref{f2}) and (\ref{peq}).

2) The spectrum of the downstream CR electrons that have cooled for the advection time, $t=d/u_2$, can be approximated with $g_e(p,d) = G\left[p/{(1- d/u_2t_{\rm rad})}\right]$ at the distance $d$ from the shock location. Here, $G$ is the functional form of the spectrum at the shock location of $d = 0$. The synchrotron emission from this analytic formula provides a reasonable approximation to that calculated using DSA simulation results (see Figure 2).

3) Both the models of $M = 4.5$ shock without pre-existing CR electrons and $M = 2$ shock with pre-existing CR electrons of $f_{e,1} \propto p^{-4.2}$ may explain the observed properties of the radio relic in CIZA J2242.8+5301. The postshock electron CR number fraction of $\xi_e \sim 10^{-7}$ in the injection-only model or the ratio of upstream CR electrons pressure to gas pressure of $R_1 \sim {\rm several} \times 10^{-5}$ in the model with pre-existing CRs are required to explain the observed radio flux of several mJy. Those values of $\xi_e$ and $R_1$ are modest enough to be accommodated in typical clusters. But the surface of the shock responsible for the relic should be highly elongated with the aspect ratio of $\sim 3 - 3.5$. It would not be trivial for such structure to be induced in merger events in clusters.

4) The radio relic in ZwCl0008.8+5215 may be explained by the models of $M = 2$ shock with or without pre-existing CR electrons. However, in the injection-only model, $\xi_e \ga 10^{-4}$, required to explain the observed radio flux of several mJy, is probably too large for the weak shock of $M = 2$. On the other hand, in the model with pre-existing CRs, $R_1 \sim 10^{-3}$, required to explain the observed flux, seems to be marginal, that is, not inconsistent with expected values in clusters. In the model, then, the origin of such pre-existing electron population is an important topic, but beyond the scope of the present paper.

5) The color-color diagram of $\alpha_{91}^{20}$ vs $\alpha_{20}^{6}$ has been presented behind the projected shock edge. It includes an important information about the evolutionary properties of the postshock electrons. Due to the effect of the projection with limited subtended angle along the LoS for spherical shocks, the diagram behaves differently for different projection angles. So it may provide an independent way to estimate the projection angle, which is a key parameter in modeling of radio relics.

\acknowledgements

HK was supported by Basic Science Research Program through the National Research Foundation of Korea funded by the Ministry of Education, Science and Technology (2011-0002433). DR was supported by the National Research Foundation of Korea through grant 2007-0093860. TWJ was supported by NASA grant NNX09AH78G, NSF grant AST-0908668 and by the Minnesota Supercomputing Institute for Advanced Computational Research. We thanks R. J. van Weeren and L. Rudnick for discussions.

\clearpage

\begin{deluxetable} {cccccccc}
\tablecaption{Parameters for Plane-Parallel Shock Simulations}
\tablehead{
\colhead {Model Name} & \colhead{$z$} & \colhead{$c_{s,1}$ } & \colhead{$M$} 
& \colhead{$u_2$} & \colhead{$s$} & \colhead{$B_2$} & \colhead{Cluster} \\
\colhead { } & \colhead { } & \colhead {(${\rm km~s^{-1}}$)} & \colhead { } 
 & \colhead {(${\rm km~s^{-1}}$)} & \colhead { } & \colhead {($\mu$G) } & \colhead { }
}
\startdata
M4.5B7I & 0.1921 & $7.8\times10^2$ & 4.5 &$1.0\times10^3$ & - & 7.0 & CIZA J2242.8+5301\\
M4.5B3.5I & 0.1921 & $6.0\times10^2$ & 4.5 &$7.7\times10^2$ & - & 3.5 & CIZA J2242.8+5301\\
M2B2.3S4.2 & 0.1921 & $1.25\times10^3$ & 2.0 &$1.1\times10^3$ & 4.2 & 2.3 & CIZA J2242.8+5301\\
M2B7S4.2 & 0.1921 & $1.25\times10^3$ & 2.0 &$1.1\times10^3$ & 4.2 & 7.0 & CIZA J2242.8+5301\\
M2B2.3I & 0.103 & $1.25\times10^3$ & 2.0 &$1.1\times10^3$ & - & 2.3 & ZwCl0008.8+5215\\
M2B2.3S5.4 & 0.103 &$1.25\times10^3$ & 2.0 &$1.1\times10^3$ & 5.4 & 2.3 & ZwCl0008.8+5215\\
\enddata
\end{deluxetable}

\clearpage

\begin{figure}
\vspace{-0.5cm}
\hskip -0.8cm
\includegraphics[scale=0.87]{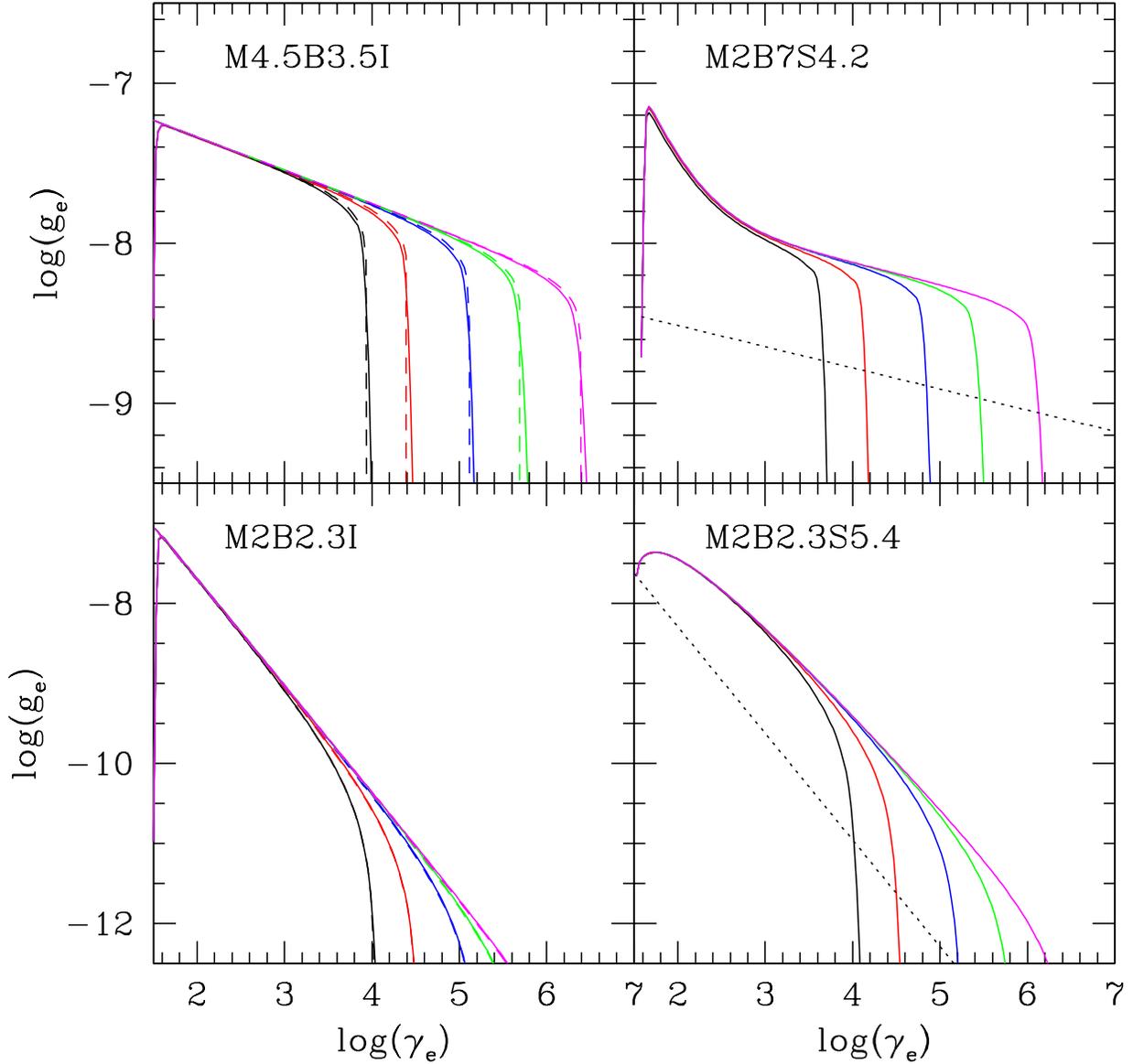}
\vspace{-1.5cm}
\caption{The distribution function, $g_e=f_{e,2}(\gamma_e)\gamma_e^4$, of CR electrons at 5 different locations downstream of the shock for the M4.5B3.5I, M2B7S4.2, M2B2.3I, and M2B2.3S5.4 models (see Table 1 for model parameters). In this figure the postshock CR electron number fraction is set to be $\xi_e = 10^{-6}$ for comparison of the models. In all the models, solid lines are the results of DSA simulations. In M4.5B3.5I and M2B2.3I, dashed lines show the approximate solution given in equation (\ref{gapprox}). In M2B7S4.2 and M2B2.3S5.4, the distribution of the pre-existing population of CR electrons is shown with dotted lines. The five downstream locations are: $d=0.30$, 1.4, 5.8, 29, 86 kpc in M4.5B3.5I, $d=0.34$, 1.7, 6.l, 34, 100 kpc in M2B7S4.2, $d=0.42$, 2.0, 8.1, 40, 120 kpc in M2B2.3I and M2B2.3S4.2.}
\label{Fig1}
\end{figure}

\clearpage

\begin{figure}
\vspace{-1cm}
\hskip -0.8cm
\includegraphics[scale=0.87]{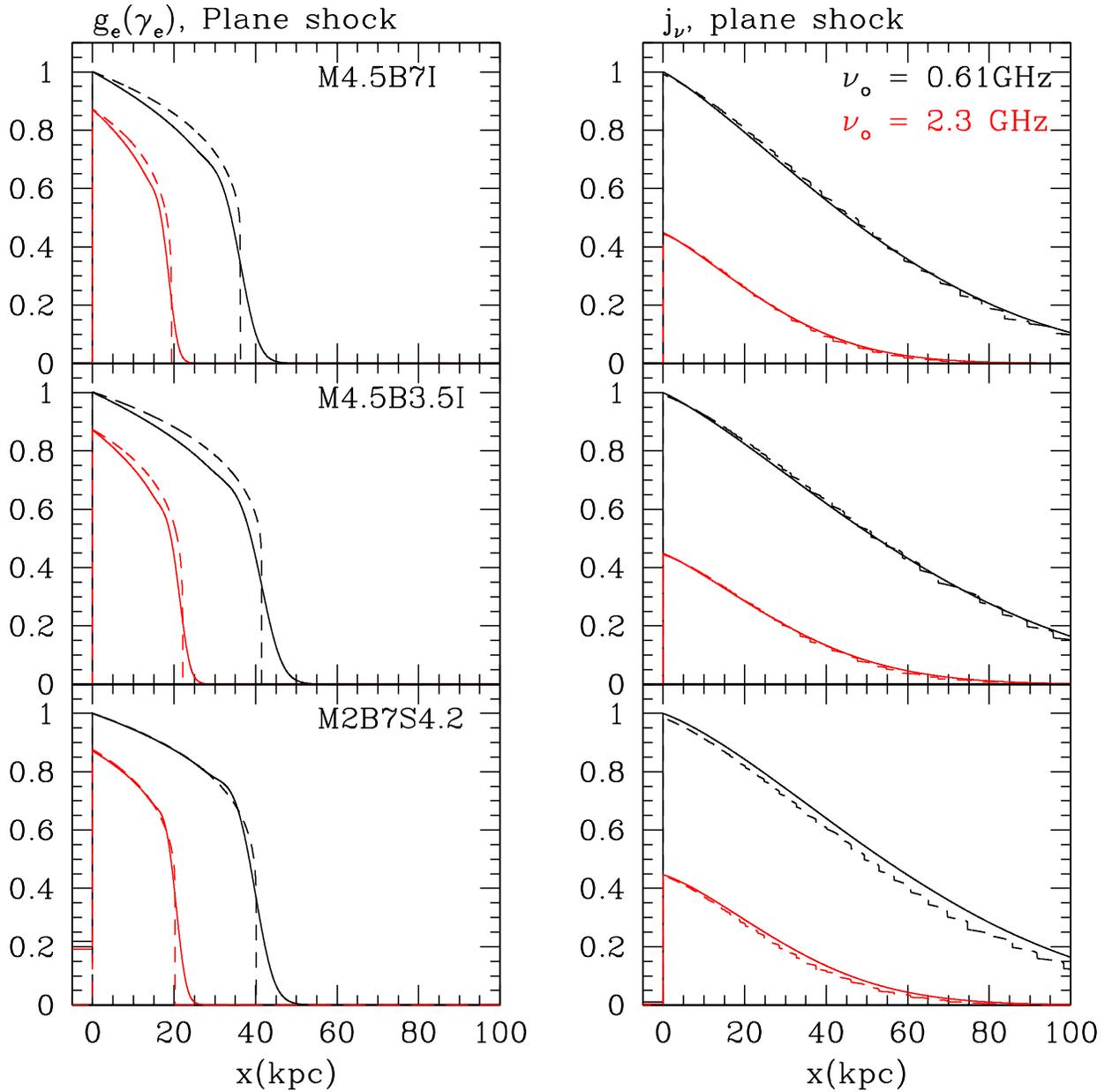}
\vspace{-0.5cm}
\caption{ Left: The distribution function, $g_e(x)$, of CR electrons with $\gamma_e$'s given in equation (\ref{fpeak}) for $\nu_{\rm obs} =$ 0.61 and 2.3 GHz in the downstream region for the M4.5B7I, M4.5B3.5I, and M2B7S4.2 models. Right: The synchrotron emissivity, $j_{\nu}(x)$, at $\nu_{\rm obs} =$ 0.61 and 2.3 GHz for the same three models. Solid lines are the results of DSA simulations, and dashed lines show the approximate solution given in equation (\ref{gapprox}) and the emissivity calculated using the solution.}
\label{Fig2}
\end{figure}

\clearpage

\begin{figure}
\vspace{-1cm}
\hskip -0.8cm
\includegraphics[scale=0.87]{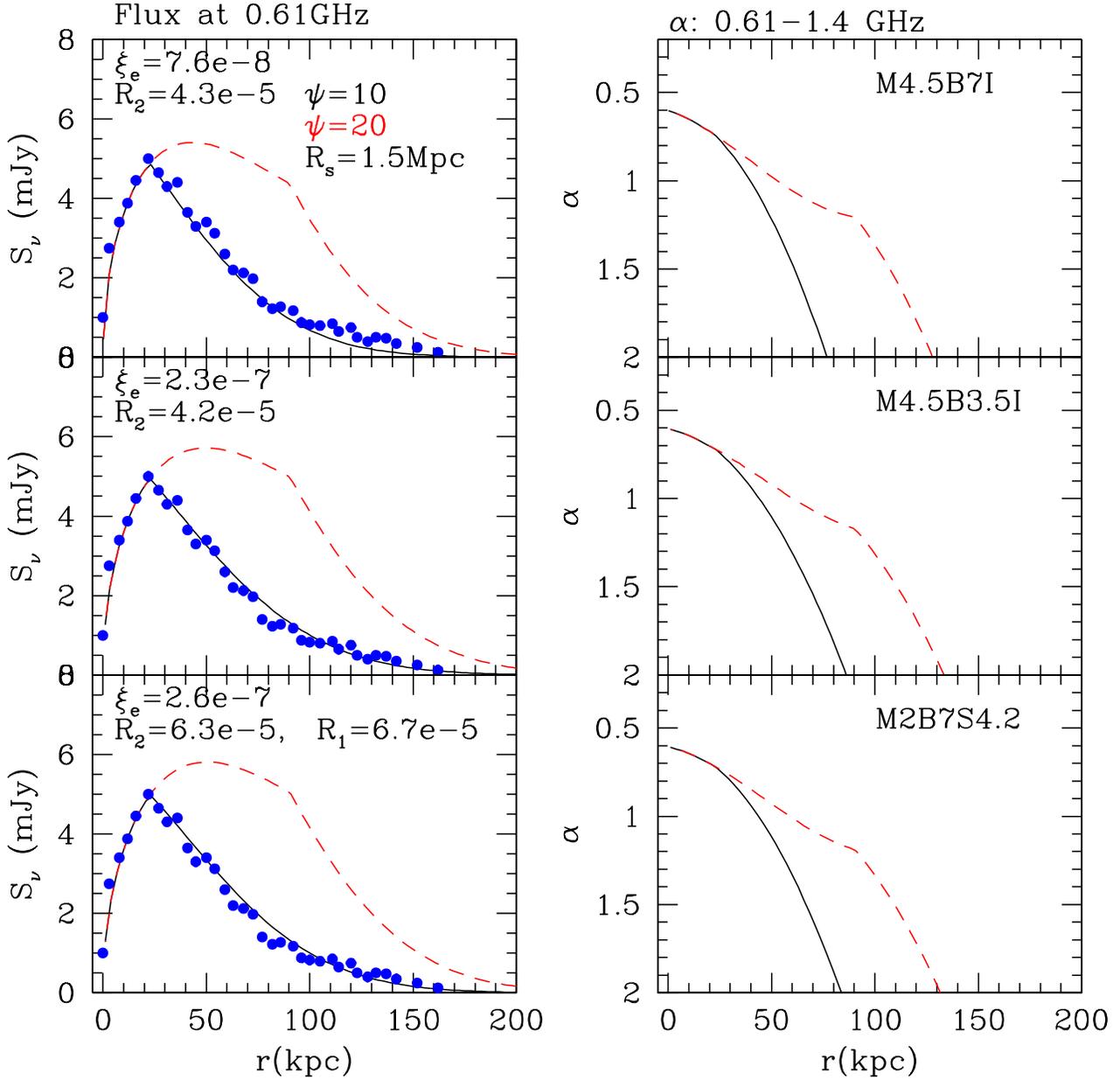}
\vspace{-0.5cm}
\caption{The synchrotron flux, $S_{\nu}$, at 0.61GHz and the spectral index, $\alpha$, between 0.61 GHz and 1.4 GHz for the M4.5B7I, M4.5B3.5I, and M2B7S4.2 models. Spherical shocks with radius $R_s=1.5$ Mpc are assumed, and two projection angles, $\psi=10^{\circ}$ (solid lines) and $20^{\circ}$ (dashed lines), are considered. Filled circles are the data points taken from \citet{vanweeren10} for the radio relic in CIZA J2242.8+5301. The flux is scaled so that the peak has 5 mJy. The required values of the postshock electron CR number fraction, $\xi_e$, and the ratio of upstream and downstream CR electrons pressure to gas pressure, $R_1$ and $R_2$, are shown.}
\label{Fig3}
\end{figure}

\clearpage

\begin{figure}
\vspace{1cm}
\hskip -0.8cm
\includegraphics[scale=0.87]{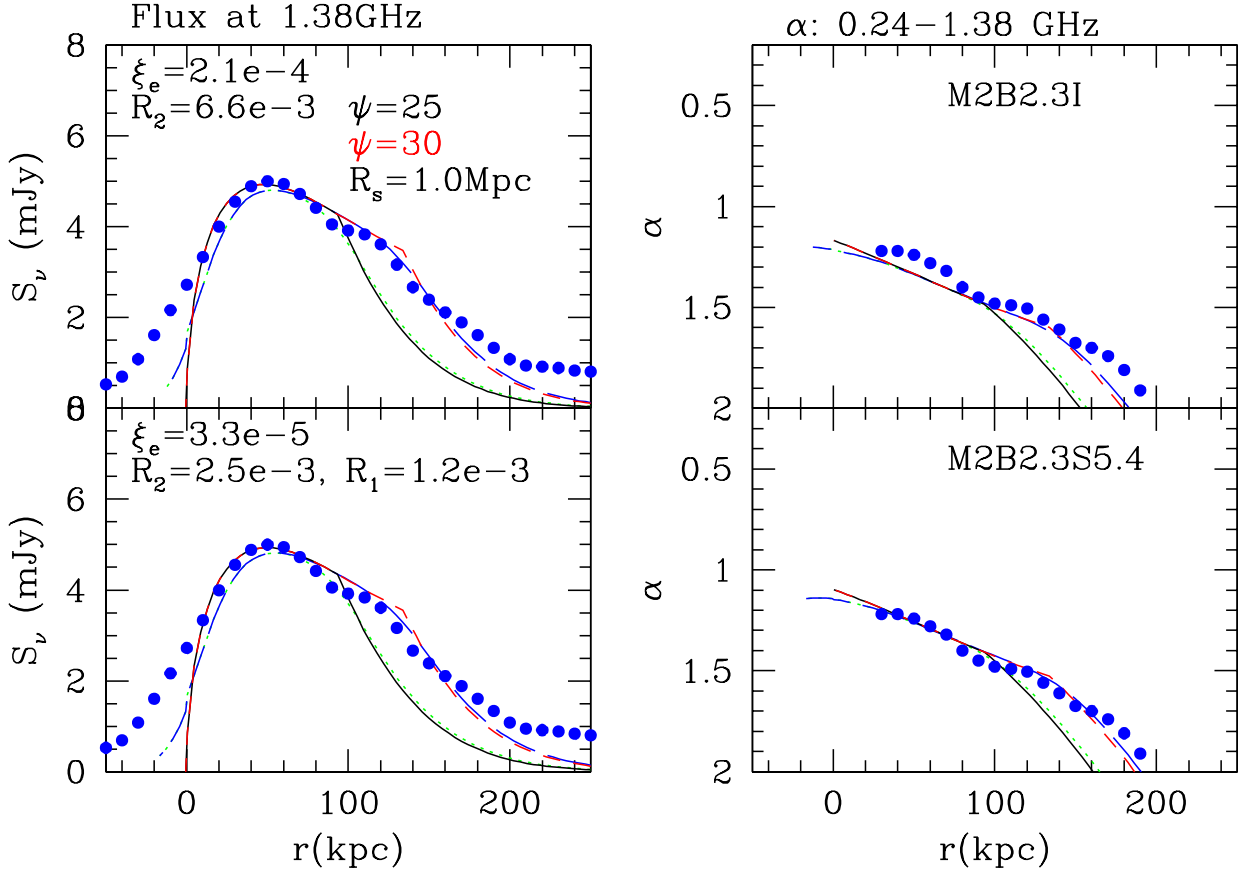}
\vspace{-5.5cm}
\caption{The synchrotron flux, $S_{\nu}$, at 1.38GHz and the spectral index, $\alpha$, between 0.24 GHz and 1.38 GHz for the M2B2.3I and M2B2.3S5.4 models. Spherical shocks with radius $R_s=1.0$ Mpc are assumed, and two projection angles, $\psi=25^{\circ}$ (solid lines) and $30^{\circ}$ (dashed lines), are considered.
Dotted and long-dashed lines are the results convolved with a Gausian beam with the e-width of $12^"$.
Filled circles are the data points taken from \citet{vanweeren11} for the radio relic in ZwCl0008.8+5215. The flux is scaled so that the peak has 5 mJy. The required values of the postshock electron CR number fraction, $\xi_e$, and the ratio of upstream and downstream CR electrons pressure to gas pressure, $R_1$ and $R_2$, are shown.}
\label{Fig4}
\end{figure}

\clearpage

\begin{figure}
\vspace{-1cm}
\hskip -0.5cm
\includegraphics[scale=0.87]{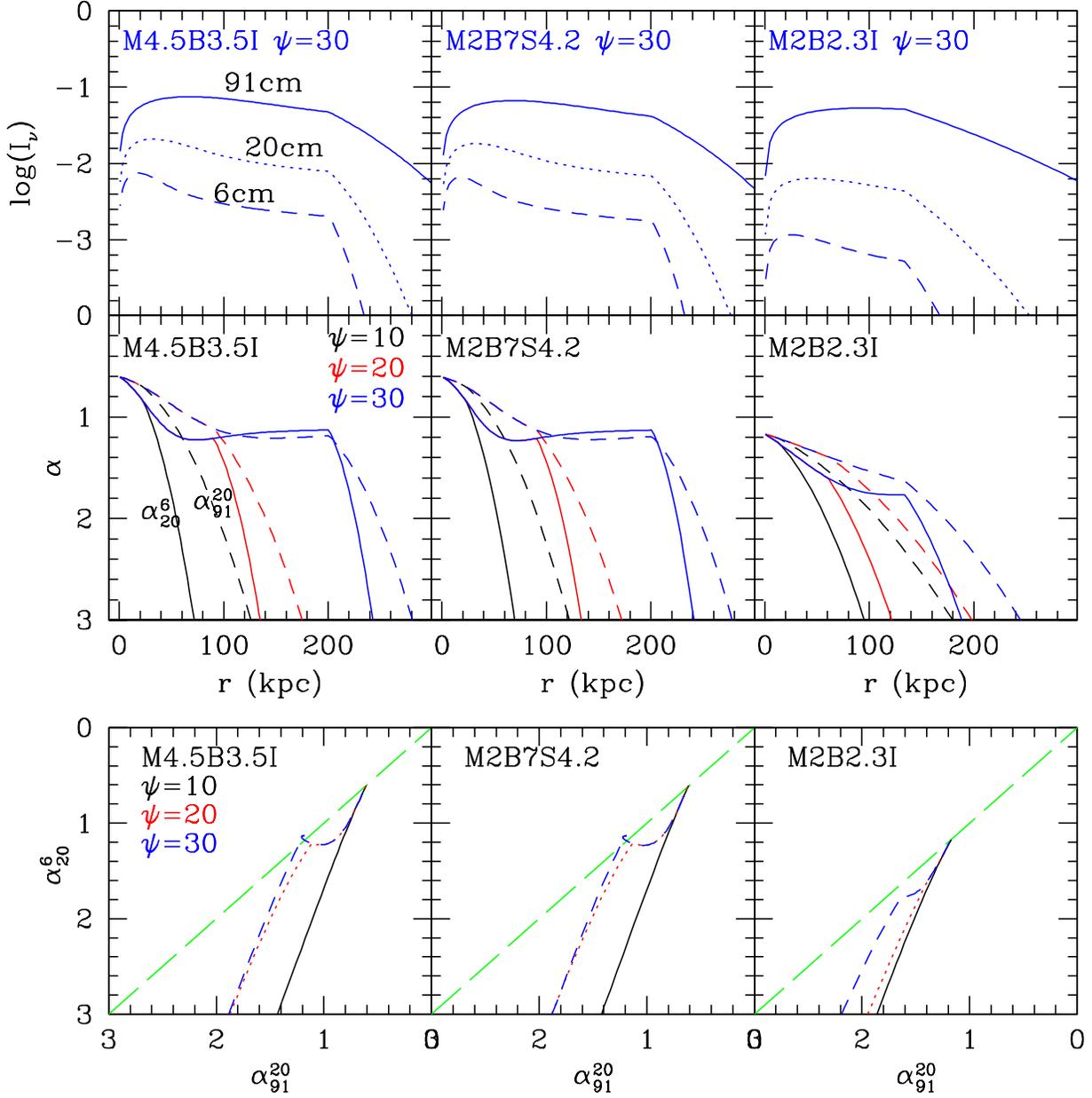}
\vspace{-0.5cm}
\caption{Top: The intensities at 6 cm (dashed lines), 20 cm (dotted), and 91 cm (solid) in the case of $\psi=30^{\circ}$ for the M4.5B3.5I, M2B7S4.2 and M2B24I models. Middle: The spectral indices $\alpha_{20}^6$ (solid lines) calculated between 6 cm and 20 cm, and $\alpha_{91}^{20}$ (dashed lines) between 20 cm and 91 cm for three different values of the projection angle, $\psi$. Bottom: The color-color diagrams of $\alpha_{91}^{20}$ vs $\alpha_{20}^{6}$ for three different values of $\psi$.}
\label{Fig5}
\end{figure}

\end{document}